\documentclass[a4paper,fleqn]{cas-sc}
\usepackage{soul}
\usepackage[sort&compress,numbers]{natbib}
\usepackage{lineno}
\usepackage{setspace}
\usepackage{multirow}

\usepackage{cuted}

\def\tsc#1{\csdef{#1}{\textsc{\lowercase{#1}}\xspace}}
\tsc{WGM}
\tsc{QE}
\tsc{EP}
\tsc{PMS}
\tsc{BEC}
\tsc{DE}
\begin{document}
\let\WriteBookmarks\relax
\def\floatpagepagefraction{1}
\def\textpagefraction{.001}
\shorttitle{Fracture Mechanism of Popgraphene Membranes}
\shortauthors{Pereira J\'unior \textit{et~al}.}

\title [mode = title]{Temperature Effects on the Fracture Dynamics and Elastic Properties of Popgraphene Membranes}

\author[1]{Marcelo L. Pereira J\'unior}
\author[1,2]{ Luiz A. Ribeiro J\'unior}
\cormark[1]
\ead{ribeirojr@unb.br}
\author[3]{ Wjefferson H. S. Brand\~ao}
\author[3]{ Acrisio L. Aguiar}
\author[4,5]{ Douglas S. Galvao}
\author[2,6]{ Jos\'e M. De Sousa}

\address[1]{Institute of Physics, University of Bras\'ilia, Bras\'ilia, 70910-900, Brazil.}
\address[2]{PPGCIMA, Campus Planaltina, University of Bras\'{i}lia, 73.345-010, Bras\'{i}lia, Brazil.}
\address[3]{Department of Physics, Federal University of Piau\'i, Teresina, Piau\'i, Brazil.}
\address[4]{Applied Physics Department, University of Campinas, Campinas, S\~ao Paulo, Brazil.}
\address[5]{Center for Computing in Engineering and Sciences, University of Campinas, Campinas, S\~ao Paulo, Brazil.}
\address[6]{Federal Institute of Education, Science and Technology of Piau\'i, S\~ao Raimundo Nonato, Piau\'i, Brazil.}

\cortext[cor1]{Corresponding author}

\begin{abstract}
Popgraphene (PopG) is a new 2D planar carbon allotrope which is composed of $5-8-5$ carbon rings. PopG is intrinsically metallic and possesses excellent thermal and mechanical stability. In this work, we report a detailed study of the thermal effects on the mechanical properties of PopG membranes using fully-atomistic reactive (ReaxFF) molecular dynamics simulations. Our results showed that PopG presents very distinct fracture mechanisms depending on the temperature and direction of the applied stretching. The main fracture dynamics trends are temperature independent and exhibit an abrupt rupture followed by fast crack propagation. The reason for this anisotropy is due to the fact that y-direction stretching leads to a deformation in the shape of the rings that cause the breaking of bonds in the pentagon-octagon and pentagon-pentagon ring connections, which is not observed for the x-direction. PopG is less stiff than graphene membranes, but the Young's modulus value is only 15\% smaller.
\end{abstract}

%\begin{graphicalabstract}
%\includegraphics[width=\linewidth]{graphical_abstract}
%\end{graphicalabstract}

%\begin{highlights}
%\item Popgraphene presents very distinct fracture mechanism depending on the temperature regime considered;
%\item The tensile resilience of popgraphene differs among the plane directions substantially;
%\item For a parallel strain, the stress for popgraphene behaves similarly to graphene at room temperature.
%\end{highlights}

\begin{keywords}
Popgraphene \sep Carbon Allotrope \sep Nanostructures \sep Mechanical/Thermal properties
\end{keywords}

%\linenumbers
\maketitle
\doublespacing

\section{Introduction}
\label{sec1}
Since graphene was obtained in 2004 \cite{novoselov2004electric}, an enormous amount of studies has been carried out aimed at using this material to develop more efficient optoelectronic devices. Among these devices, graphene-based thin-film transistors \cite{li2008graphene} and photo-detectors \cite{kim2009large} have been considered the most promising solutions to replace the current silicon-based technology, due to their good cost-efficiency ratio and possible lower environmental impact. Motivated by graphene success, several other allotropes such as penta-graphene \cite{zhang2015penta}, phagraphene \cite{wang2015phagraphene}, $\psi$-graphene \cite{li2017psi}, twin graphene \cite{jiang2017twin}, and popgraphene (PopG) \cite{wang2018popgraphene} have been proposed. These new structures exhibit several interesting physical and chemical properties that could be exploited in future applications. 

PopG is a 2D planar carbon allotrope composed of $5-8-5$ carbon rings, which was theoretically predicted by Wang and colleagues \cite{wang2018popgraphene} in 2018. PopG is particularly attractive due to its metallic behavior and excellent thermal and mechanical stabilities \cite{wang2018popgraphene,meng_PCCP}.  Also, based on density functional theory (DFT) calculations \cite{wang2018popgraphene}, it was demonstrated that PopG possesses a high capability for Lithium adsorption, as well as good conductivity and a low-energy barrier to Lithium diffusion, which makes PopG a promising material for developing Li-ion batteries. 

PopG mechanical behavior was investigated using reactive molecular dynamics simulations \cite{meng_PCCP}. They investigated the fracture of defective PopG membranes. It was concluded that the fracture patterns depend on the geometry of the defect and the involved bonds where the fracture initiates. However, a more comprehensive study of the PopG mechanical properties, in particular, its mechanical behavior dependence on thermal effects is still missing and it is one of the objectives of the present work.

In this work, the fracture patterns/mechanisms of PopG membranes subjected to different thermal regimes wer investigated using fully-atomistic reactive molecular dynamics simulations. The mechanical properties of these systems were studied using the stress-strain relationship and fracture toughness. The direction-dependence of their tensile strength is also discussed in detail. For comparison purposes, we have also carried out the same set of simulations for graphene sheets of similar dimensions. We believe that our results provide a better framework for understanding PopG mechanical behavior.  

\begin{figure}
    \centering
    \includegraphics[width=0.9\linewidth]{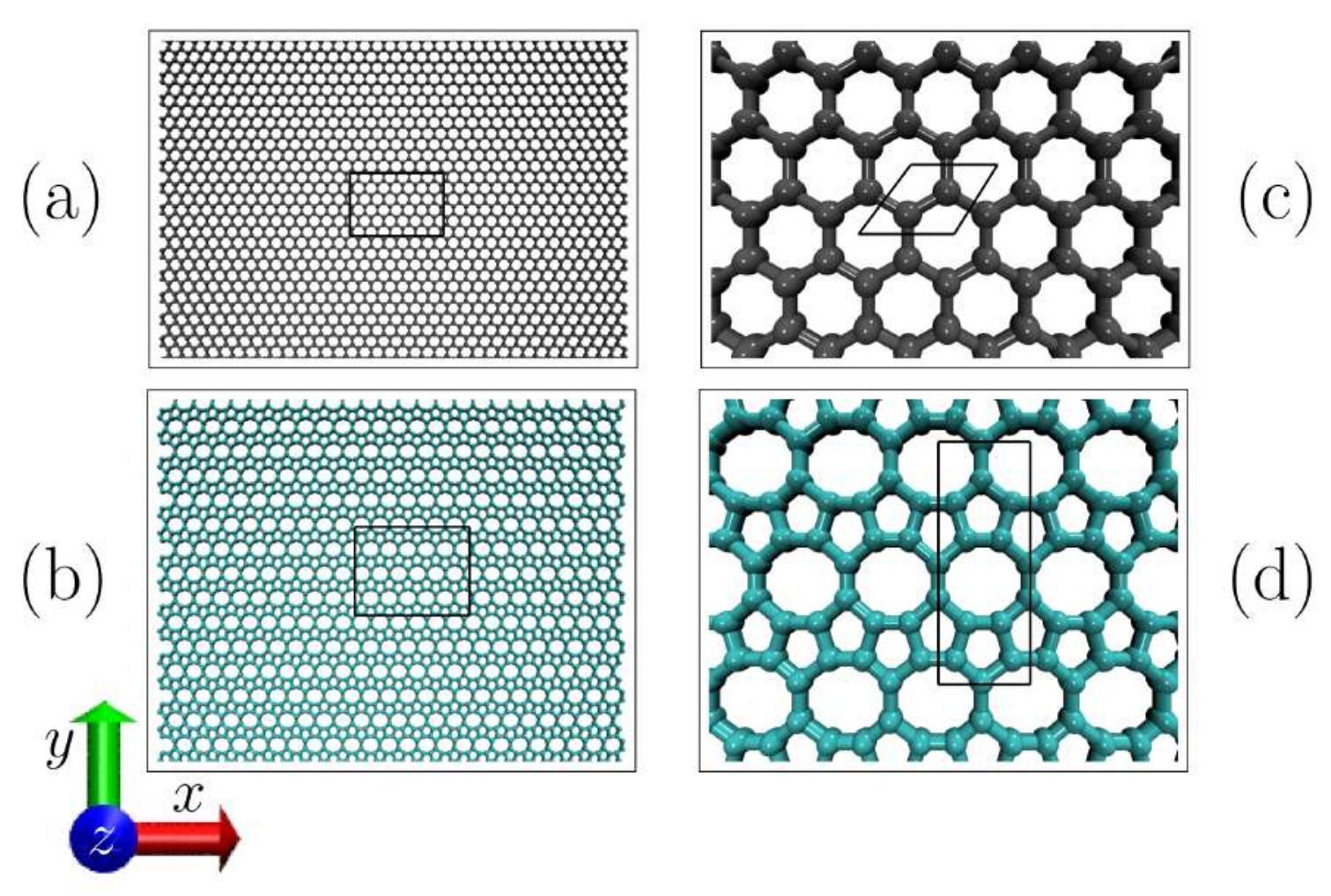}
    \caption{Schematic representation of the atomic structure of (a) graphene and (b) Popgraphene. The highlighted squares are zoomed out in panels (c) and (d), where the unit cells are also highlighted.}
    \label{fig1}
\end{figure}

\section{Computational Methods}
\label{sec2}

The unit cell of a PopG monolayer has a rectangular shape of dimensions $3.68$ \AA~by $9.11$ \AA, as illustrated in Figure \ref{fig1}. In order to study the PopG mechanical properties, we considered a supercell membrane with dimensions $109.41$ \AA~$\times$ $120.96$ \AA~containing 4752 Carbon atoms. The corresponding graphene membrane (for comparative purposes), with dimensions of $94.00$ \AA~$\times$ $91.00$ \AA, and contains 3256 atoms. To investigate the temperature effects on these membranes, we carried out fully-atomistic molecular dynamics (MD) simulations using the interatomic reactive potential ReaxFF \cite{mueller2010development,van2001reaxff} as implemented by the large-scale atomic/molecular massively parallel simulator (LAMMPS) \cite{plimpton1995fast}.

Before applying the uniaxial stretching (we considered x and y-directions), the PopG and graphene membranes were thermalized in an NPT ensemble \cite{evans1983isothermal} at 300 K and null pressure to eliminate any residual stress, during 100 ps. Starting from the thermalized structures, the strain rate is simulated by linearly increasing the simulation box along one of its periodic directions using an NVT ensemble \cite{plimpton1995fast} at 300, 600, 900, and 1200 K. In all simulations the used time step was of 0.05 fs and the temperatures were controlled using a Nos\'e-Hoover thermostat \cite{hoover1985canonical}. 

 The stress-strain curves were obtained considering an engineering strain rate of $10^{-6} fs^{-1}$. From these curves we can estimate some magnitudes such as Young’s modulus ($Y_M$), fracture Strain (FS), ultimate strength (US), and critical strain (CS). One helpful magnitude to analyze the fracture dynamics is the von Mises stress (vMS) \cite{mises_1913,de2016mechanical,de2018mechanical,de2018mechanicalNanotube,de2015hydrogenation,de2019elastic,de2017mechanical}. vMS provides local information on the accumulated stress and helps to visualize the starting fracture regions. vMS per atom $k$ is defined as:
\begin{equation}
    \sigma^{k}_{v} = \sqrt{\frac{(\sigma^{k}_{xx} - \sigma^{k}_{yy})^2 + (\sigma^{k}_{yy} - \sigma^{k}_{zz})^2 + (\sigma^{k}_{xx} - \sigma^{k}_{zz})^2 + 6((\sigma^k_{xy})^2+(\sigma^k_{yz})^2+(\sigma^k_{zx})^2)}{2}},
\end{equation}
where $\sigma^k_{xx}$, $\sigma^k_{yy}$, and $\sigma^k_{zz}$ are the components of the normal stress and $\sigma^k_{xy}$, $\sigma^k_{yz}$, and $\sigma^k_{zx}$ are the components of the shear stress. The vMS stress values are used here for better interpretation of MD snapshots and trajectories, that were obtained using free visualization and analysis software VMD \cite{HUMPHREY199633}. 

\section{Results}
\begin{figure}[pos=ht]
    \centering
    \includegraphics[width=0.7\linewidth]{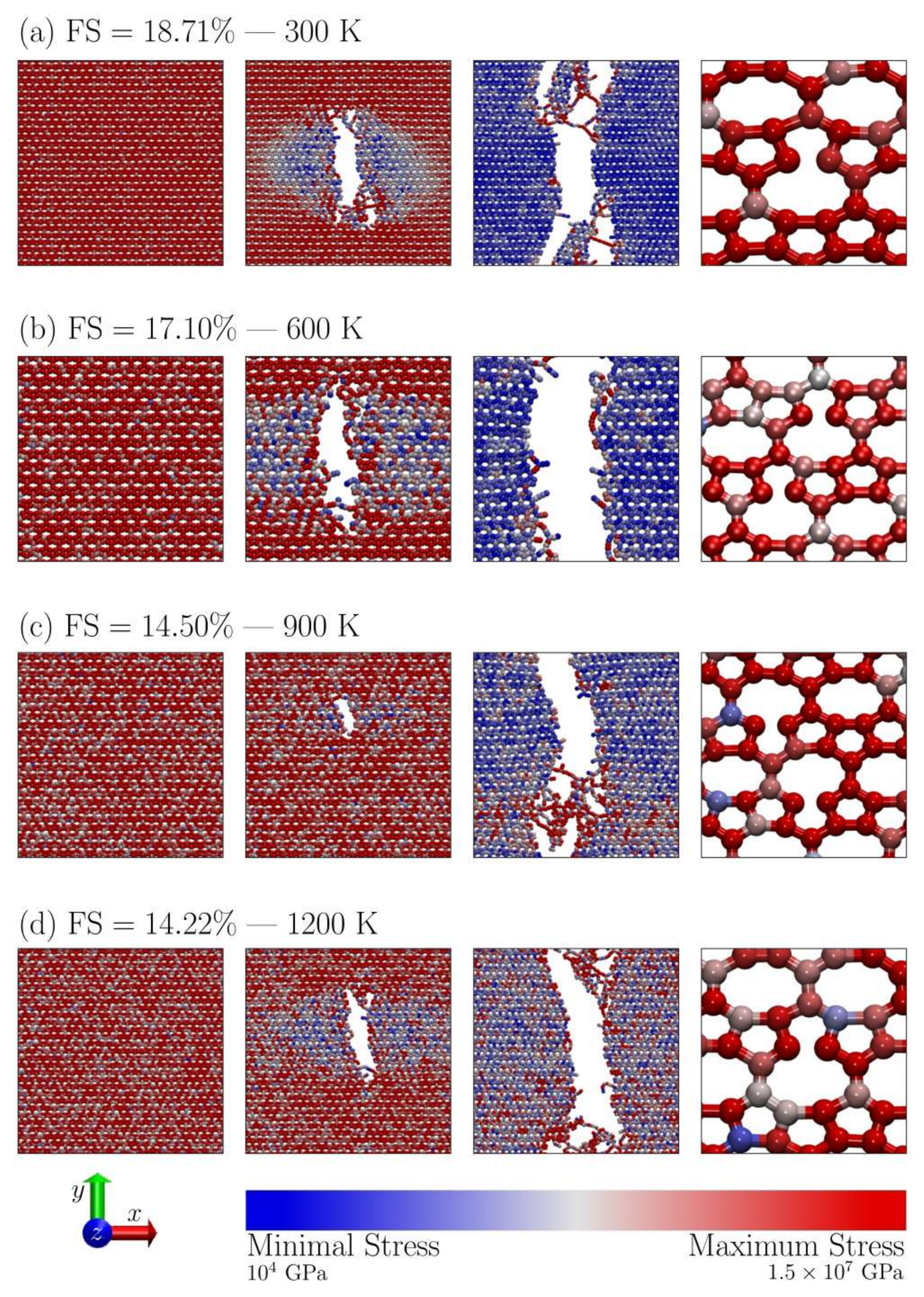}
    \caption{Representative MD snapshots from the uniaxial tensile stretching simulations (x-direction) for a PopG membrane at: (a) 300K, (b) 600 K, (c) 900 K, and (d) 1200 K. In the color scheme, blue and red represent, respectively, low and high von Mises stress values. The left-most panels show MD snapshots for just before the first fracture evidence. The two panels in the middle are snapshots at later fracture stages and crack propagation, and the right-most panels illustrate the regions from which the crack propagation has originated. The fracture strain (FS) value for each case is presented on top of their respective sequence of MD snapshots.}
    \label{fig2}
\end{figure}

We begin our discussions by presenting the fracture patterns of a PopG membrane at different temperatures and stretched along the x-direction. Figure \ref{fig2} shows a sequence of MD snapshots that starts from the one immediately before the fracture occurs (left panel in Figures \ref{fig2}(a-d)). The two panels in the middle present the snapshots of later fracture stages and crack propagation, and the right-most panels illustrate the lattice regions from which the crack propagation has originated. 

The effect of temperature on the fracture dynamics can be better understood by analyzing the local accumulations of vMS throughout the membrane, which is indicated by the color scheme in Figure \ref{fig2}, where blue and red represent the minimum and maximum stress values, respectively. As expected, we observed that increasing the temperature, more brittle becomes the PopG membrane, which is evidenced by the corresponding decrease in the FS values (from 300 K to 1200 K, FS decrease from 18.71\% to 14.22\%, respectively). However, regardless of the considered temperature values, we observed a common fracture dynamics that is an abrupt structural failure followed by fast crack propagation (the whole process is better evidenced from the videos 1 and 2 in the Supplementary Materials). 

For low temperatures before the complete structural failure, we observed minimally stressed regions (gray and blue colors), as can be seen in Figures \ref{fig2}(a-b) and video 1. However, for high temperatures (see Figures \ref{fig2}(c-d)), there is a predominance of red color regions associated with highly stressed carbon bonds that contribute to decreasing the critical strain value for the membrane fracture. After the membrane total fracture, occurs the predominance of grey and blue regions, with the thermal effects contributing to eliminate the residual stress accumulated in the bond lengths. The last column of panels in Figure \ref{fig2} zoomed-in the regions that have triggered crack propagation. From these panels, one can notice that the bond breaking mechanism involves just a bond between pentagonal and octagonal rings for all cases. Interestingly, when PopG is stretched along the x-direction, the potential energy gains are stored into all the C--C bonds without any preferred direction. This feature differs from graphene membranes, in which the straining process leads to an accumulation of potential energy in C--C bonds parallel to the stretching direction \cite{C4NR00423J}. Another important result revealed in Figure \ref{fig2} is that the PopG stretching along the x-direction tends to preserve the rings forms, which can make PopG as resilient as graphene, as discussed below. In Figures \ref{fig2}(a), \ref{fig2}(c) and \ref{fig2}(d) one can observe the appearance of linear atomic carbon chains (LACs) on the last MD snapshots. These LACs do not significantly contribute to increasing PopG membranes elasticity and are frequently observed in the fracture of low-dimensional carbon-based structures \cite{chuvilin_2009_NJP,wang_2009_PRB}. 

\begin{figure}[pos=ht]
    \centering
    \includegraphics[width=0.7\linewidth]{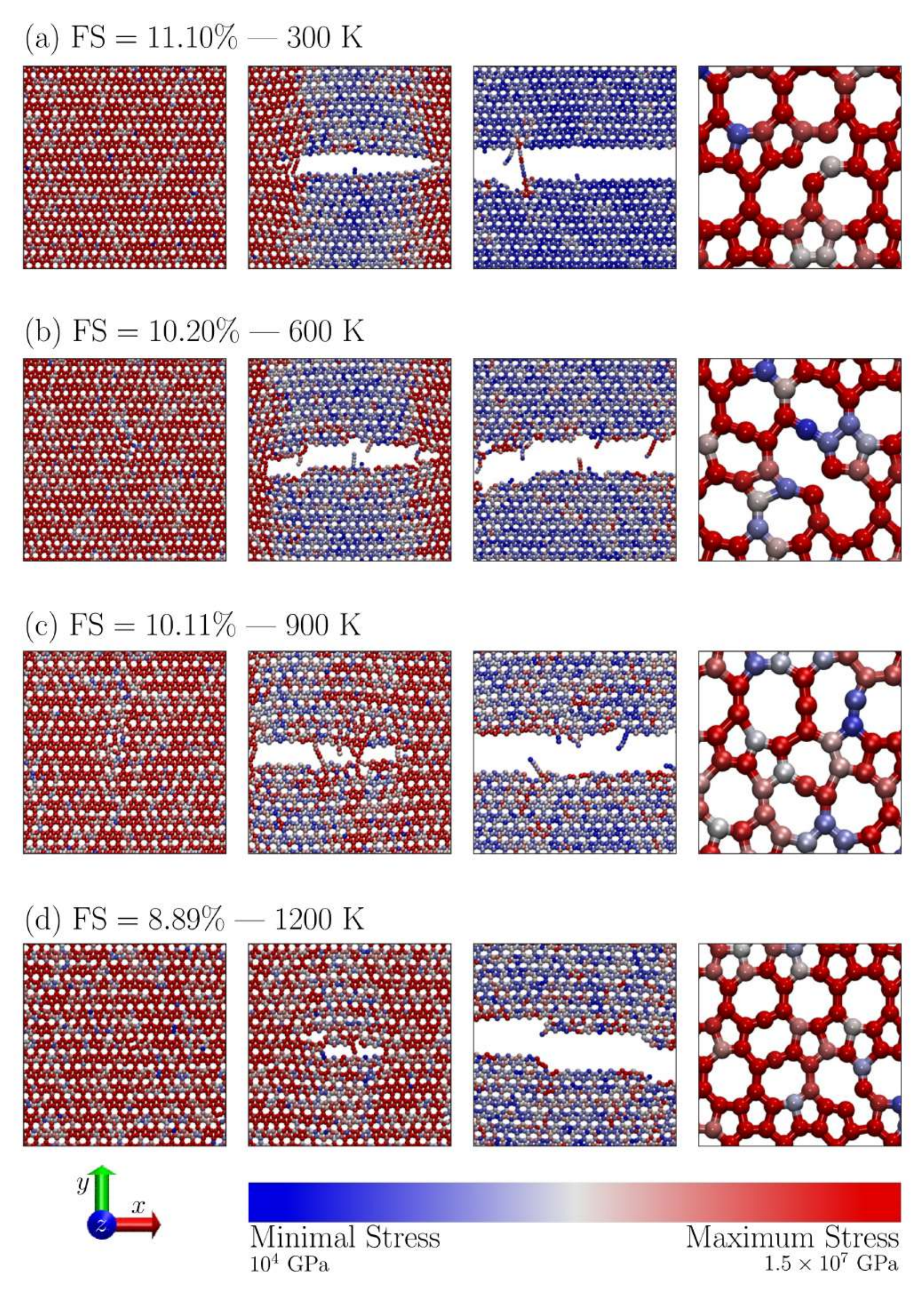}
    \caption{Representative MD snapshots from the uniaxial tensile stretching simulations (y-direction) for a PopG membrane at: (a) 300K, (b) 600 K, (c) 900 K, and (d) 1200 K. In the color scheme, blue and red represent, respectively, low and high von Mises stress values. The left-most panels show MD snapshots for just before the first fracture evidence. The two panels in the middle are snapshots at later fracture stages and crack propagation, and the right-most panels illustrate the regions from which the crack propagation has originated. The fracture strain (FS) value for each case is presented on top of their respective sequence of MD snapshots.}
    \label{fig3}
\end{figure}

In Figure \ref{fig3} we present the corresponding results for the strain applied along the y-direction, considering the same temperature regimes for the case discussed above (x-direction stretching). Remarkably, the results in Figure \ref{fig3} and video 2 reveal that PopG membranes present a substantial degree of fracture anisotropy in relation to the direction of the applied stretching, for the x and y-directions the fracture lines are mainly vertical and horizontal, respectively. This behavior is different from the graphene case, as it will be discussed below. The critical strain that initiates the fracture processes ranges from 11.10\% to 8.89\% for temperature values varying from 300 K up to 1200 K, respectively. These FS values are substantially smaller than the ones obtained for the x-direction stretching (see Figure \ref{fig2}). The y-direction stretching leads to a deformation in the shape of the rings that causes the breaking of bonds in the pentagon-octagon and pentagon-pentagon ring connections, as illustrated in the left-most panels of Figure \ref{fig3}. These results suggest that the dynamical deformation of both pentagonal and octagonal rings, during the y-direction stretching, is the main reason for the observed degree of anisotropy presented in the PopG straining process. In contrast to the cases presented in Figure \ref{fig2}, we can see the formation of very small LACs after the complete rupture of the PopG membrane. This abrupt rupture is a consequence of the ring deformation process mentioned above. As can be also inferred from Figure \ref{fig3}, temperature acts similarly to the simulations presented in Figure \ref{fig2} in structurally destabilizing the PopG membranes. 

\begin{figure}[pos=ht]
    \centering
    \includegraphics[width=0.7\linewidth]{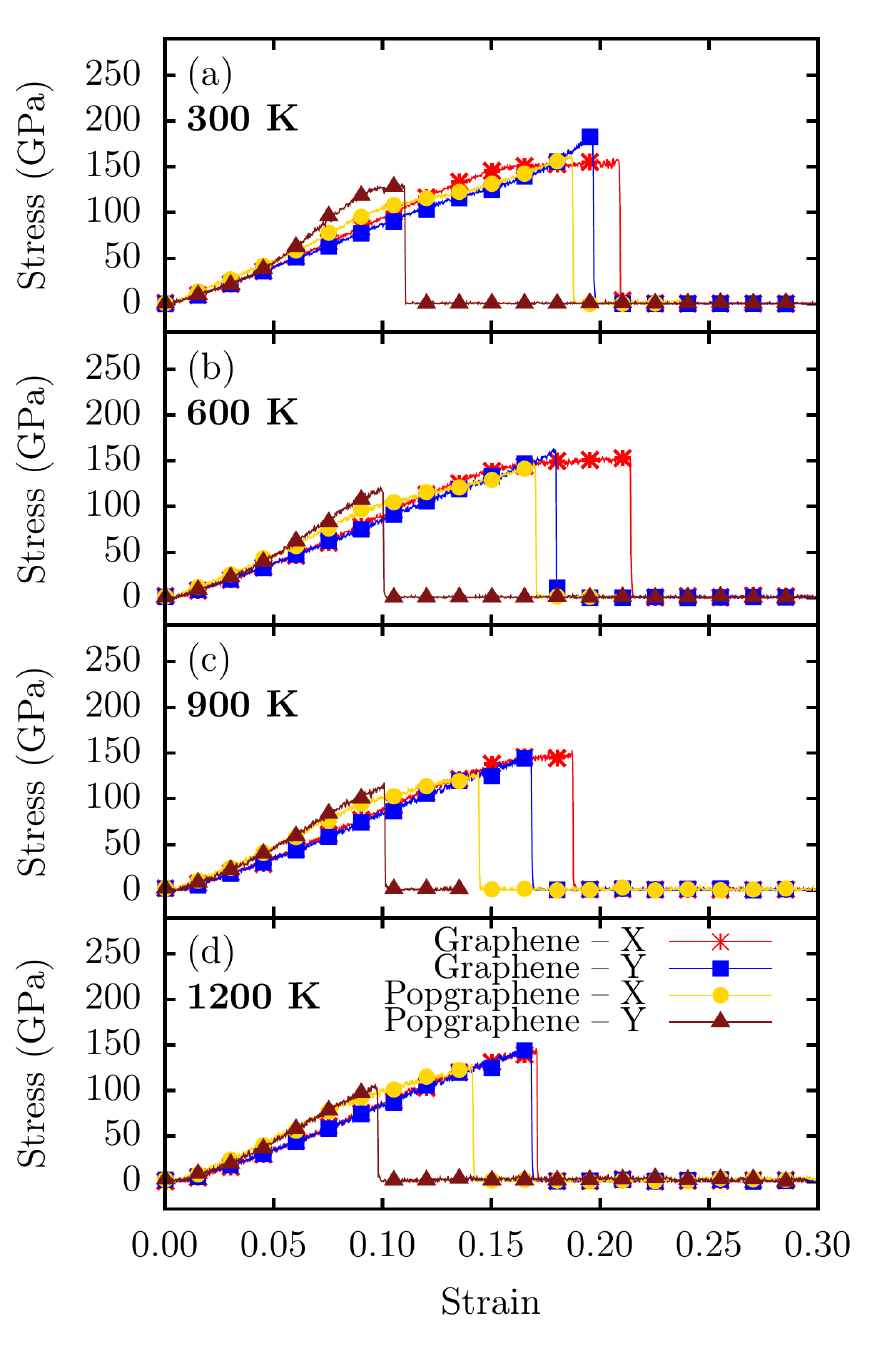}
    \caption{Stress-strain curves for PopG membranes at: (a) 300 K, (b) 600 K, (c) 900 K, and (d) 1200 K, considering the cases presented in Figures \ref{fig2} and \ref{fig3}. For comparison purposes, the stress-strain curves for graphene membranes are also shown. (--X) and (--Y) refer to the stretching directions.}
    \label{fig4}
\end{figure}

In Figure \ref{fig4} we present the calculated stress-strain curves for PopG and graphene, considering the same temperature values presented above. In this Figure, (--X) and (--Y) refer to the applied stretching directions. The stress-strain curves shown in Figure \ref{fig4} present two common and distinct regions: (I) a quasi-linear elastic regime is observed up to a critical strain value (fracture strain) and (II) a fractured regime takes place after strain values higher than the critical one, in which the stress values goes abruptly to zero. At room temperature, Figure \ref{fig4}(a), PopG-X and graphene(--X,--Y) show similar tensile strength, with a critical strain about 0.2\%. The PopG--Y curve presents a critical strain of approximately 0.12\%. As mentioned above, this difference is due to the carbon rings deformation process that occurs during the y-direction stretching. This process is not significantly temperature-dependent, as can be inferred from Figure \ref{fig4}, as the FS values are almost the same. Interestingly, the difference in the critical strength values for PopG--X and graphene--X simulations increases for the temperatures 600 K and 900 K, as shown in Figures \ref{fig4}(b) and \ref{fig4}(d), respectively.  For temperatures about 1200 K, and higher than it, PopG and graphene tend to present similar trends and the difference in the critical strain for the total rupture between PopG and graphene decreases.       

In Table \ref{table1} we present a summary of the elastic properties (Young's modulus ($Y_M$), fracture strain (FS), and the ultimate strength (US)) of graphene and PopG. These properties were derived from fitting the linear region of the stress-strain curves presented in Figure \ref{fig4}. PopG and graphene present $Y_M$ and $US$ values of the same order of magnitude, with PopG being slightly softer. The presence of pentagons and octagons makes the PopG membranes more porous and also mechanically less stiff then graphene membranes. It is worthwhile to stress here that the mechanical properties obtained here for PopG and graphene are in good agreement with previous theoretical values reported in the literature using AIREBO interatomic potential \cite{zhong_2019,anthea_AIREBO} and force field ReaxFF \cite{jensen_REAX} potentials, for graphene, and for AIREBO potential for PopG \cite{meng_PCCP}.

\begin{table}[pos=ht]
\centering
\begin{tabular}{cccccccccc}
	\hline\hline
	                         \multicolumn{4}{c}{Graphene}                           &              &  &                 &                     &              &  \\ \hline
	\multirow{2}{*}{Temperature {[}K{]}} &  &          \multicolumn{3}{c}{$x$-direction}           &  &          \multicolumn{3}{c}{$y$-direction}           &  \\ \cline{3-5}\cline{7-10}
	                                     &  & $Y_M$ {[}GPa{]} & FS [\%] & US {[}GPs{]} &  & $Y_M$ {[}GPa{]} & FS [\%] & US {[}GPs{]} &  \\ \cline{1-1}\cline{3-5}\cline{7-10}
	                300                  &  &     992.07      &        21.40        &    161.73    &  &     986.34      &        19.71        &    130.90    &  \\ \cline{1-1}\cline{3-5}\cline{7-10}
	                600                  &  &     987.29      &        20.91        &    146.35    &  &     958.65      &        17.98        &    120.10    &  \\ \cline{1-1}\cline{3-5}\cline{7-10}
	                900                  &  &     983.04      &        18.80        &    128.22    &  &     944.82      &        16.84        &    117.69    &  \\ \cline{1-1}\cline{3-5}\cline{7-10}
	                1200                 &  &     976.43      &        17.18        &    128.17    &  &     889.68      &        15.91         &    105.32    &  \\ \hline\hline
\end{tabular}
\begin{tabular}{cccccccccc}
	                        \multicolumn{4}{c}{Popgraphene}                         &              &  &                 &                     &              &  \\ \hline
	\multirow{2}{*}{Temperature {[}K{]}} &  &          \multicolumn{3}{c}{$x$-direction}           &  &          \multicolumn{3}{c}{$y$-direction}           &  \\ \cline{3-5}\cline{7-10}
	                                     &  & $Y_M$ {[}GPa{]} & FS [\%] & US {[}GPs{]} &  & $Y_M$ {[}GPa{]} & FS [\%] & US {[}GPs{]} &  \\ \cline{1-1}\cline{3-5}\cline{7-10}
	                300                  &  &     847.72      &        18.71        &    157.54    &  &     819.20      &        11.10        &    185.14    &  \\ \cline{1-1}\cline{3-5}\cline{7-10}
	                600                  &  &     813.41      &        17.10        &    155.24    &  &     807.12      &        10.20        &    161.23    &  \\ \cline{1-1}\cline{3-5}\cline{7-10}
	                900                  &  &     794.17      &        14.50        &    152.58    &  &     761.94      &        10.11        &    149.50    &  \\ \cline{1-1}\cline{3-5}\cline{7-10}
	                1200                 &  &     779.80      &        14.22        &    145.94    &  &     781.91      &        8.89        &    144.10    &  \\ \hline\hline
\end{tabular}
\caption{Elastic properties of PopG and graphene: Young's modulus ($Y_M$), fracture strain (FS), and ultimate strength (US).}
\label{table1}
\end{table}

\section{Conclusions}
In summary, we have used fully-atomistic reactive molecular dynamics simulations to investigate the elastic properties of PopG membranes at different temperatures. Our results showed that PopG presents very distinct fracture mechanisms depending on the temperature and direction of the applied stretching. The main fracture dynamics trends are temperature independent and exhibit an abrupt rupture followed by fast crack propagation. For temperature values ranging from 300 K to 1200 K, the critical strain varies from 18.71\% to 14.22\%, and from 11.10\% to 8.89\% for the x and y-directions, respectively. In contrast to graphene, PopG shows a significant anisotropy regarding tensile stretching. For the x-direction stretching, PopG has tensile resilience comparable to graphene at room temperature. The reason for this anisotropy is due to the fact that y-direction stretching leads to a deformation in the shape of the rings that cause the breaking of bonds in the pentagon-octagon and pentagon-pentagon ring connections, which is not observed for the x-direction. Importantly, the presence of pentagons and octagons make the PopG membranes more porous and also mechanically less stiff than graphene membranes, but the Young's modulus value is only 15\% smaller.

\section*{Acknowledgements}
The authors gratefully acknowledge the financial support from Brazilian research agencies CNPq, FAPESP, and FAP-DF. L.A.R.J acknowledges the financial support from a Brazilian Research Council FAP-DF and CNPq grants $00193.0000248/2019-32$ and $302236/2018-0$, respectively. J. M. S. nad D.S.G. thank the Center for Computing in Engineering and Sciences at Unicamp for financial support through the FAPESP/CEPID Grants \#2013/08293-7 and \#2018/11352-7. W.H.S.B. and A.L.A. thank the Laborat\'orio de Simula\c c\~ao Computacional Caju\'ina (LSCC) at Universidade Federal do Piau\'i for computational support. L.A.R.J., A.L.A, M.L.P.J., and W.H.S.B. acknowledges CENAPAD-SP for providing the computational facilities.

\printcredits

\bibliographystyle{unsrt}

\bibliography{cas-refs}

\end{document}